\documentclass[11pt]{article}
\usepackage{graphicx}
\usepackage{amsmath,amsfonts,amssymb,amsthm}
\usepackage{gensymb}
\usepackage{textcomp}

\title{Polarization structure of gravitational waves in extended relativity
\\ \small{Accepted for publication in Classical and Quantum Gravity (IOP Publishing)}}

\author{ Y. Friedman\\
Extended Relativity Research Center\\
Jerusalem College of Technology, Israel\\
P.O.B. 16031, Jerusalem 91160, Israel\\
e-mail: friedman@g.jct.ac.il}

\begin{document}
\date{}

\maketitle
\bigskip
\noindent
{\small{\bf ABSTRACT.}
We investigate the polarization structure of gravitational waves in the framework of Extended Relativity (ER), a Lorentz-covariant theory in which gravitational fields are described by deviations from flat spacetime defined relative to a distant inertial observer. Starting from the ER point-source solution, we derive the gravitational radiation produced by a compact binary system in the wave zone and compute the associated geodesic-deviation field to second order in the orbital velocity parameter. The analysis does not rely on a weak-field perturbation of Einstein’s equations or on imposing a transverse–traceless gauge condition.

The physical polarization content is characterized by the tidal matrix $\mathcal{E}_{ab}=R_{0a0b}$,
which determines the relative acceleration of nearby test particles and provides a gauge-invariant description of gravitational-wave polarization. We show that the transverse tensor sector reproduces the standard quadrupolar $+$ and $\times$ modes at leading order. However, the spatial trace of the deviation tensor is non-vanishing, leading to additional breathing, vector, and longitudinal tidal components. These contributions are not independent propagating degrees of freedom but are geometrically correlated with the tensor sector through the source configuration. The resulting polarization pattern provides a potential observational test distinguishing ER from General Relativity while remaining compatible with current observational bounds.

We derive the corresponding detector response and discuss observational signatures for interferometric detectors and pulsar-timing arrays. The resulting polarization pattern provides a potential observational test distinguishing ER from General Relativity.}


\maketitle

\section{Introduction}

Gravitational waves (GWs) were predicted by Einstein as a consequence of General Relativity (GR) \cite{Einstein16,Einstein18}. In the weak-field regime, gravitational radiation is commonly described in the transverse–traceless (TT) gauge, in which only two tensor polarization states, conventionally denoted $(+,\times)$, remain \cite{Thorne,Isi}. More generally, metric theories of gravity may admit up to six independent polarization modes. These can be classified according to the E(2) scheme introduced by Eardley \emph{et al.} \cite{Eardley,Eardley2} and further developed in subsequent analyses \cite{Nishizawa}. Determining the polarization content of gravitational radiation therefore provides a powerful test of the dynamical structure of gravity.

The direct detections of gravitational waves by the LIGO–Virgo–KAGRA network \cite{Abbott,Abbott17,Abbott17b} have made it possible to probe the polarization structure of gravitational radiation observationally. In particular, multi-detector observations allow one to constrain possible deviations from the purely transverse tensor radiation predicted by GR. A variety of studies have investigated how additional scalar or vector polarizations may be tested using morphology-independent analyses and network response methods \cite{Schumacher23,Schumacher25,Dong,Lai,Wu,mixed}. These analyses typically parameterize the detector response in terms of the six E(2) polarization modes, allowing for independent time-dependent amplitudes. Current observations strongly favor the presence of tensor modes, but mixed configurations containing additional components are not yet completely excluded.

Extended Relativity (ER) is a relativistic framework for gravitation introduced in \cite{Friedman22} and further developed in \cite{ana, FSuper}. In this formulation, gravitational fields are represented by a deviation tensor defined on a Minkowski background and constructed covariantly from retarded source data. In contrast to the TT formulation of GR, the formalism does not impose gauge conditions that eliminate trace or longitudinal components at the outset. Consequently, the polarization content of gravitational radiation must be determined directly from the physical tidal field generated by the source.

In this work we investigate the gravitational radiation produced by a compact binary system within the ER framework. Starting from the ER point-source solution, we derive the deviation tensor for a binary in circular motion in the wave zone, retaining all contributions up to second order in the orbital velocity parameter $\beta=v/c$. The calculation consistently incorporates the Lorentz factor of the source four-velocity and the first-order retardation effects associated with the motion of the source.

The observable polarization content is obtained from the physical tidal matrix $\mathcal{E}_{ab}=R_{0a0b}$,
which governs the relative acceleration of nearby freely falling test particles and provides a gauge-invariant characterization of gravitational-wave polarization.The components of the tidal matrix can be mapped onto the standard E(2) polarization basis, allowing a direct comparison with the polarization modes used in gravitational-wave data analysis. We show that the transverse $(+,\times)$ tensor sector reproduces the familiar quadrupole pattern at leading order. However, the spatial trace of the deviation tensor does not vanish, leading to additional breathing, vector-like, and longitudinal tidal components. These contributions are not independent propagating degrees of freedom but are correlated with the tensor sector through the geometry and dynamics of the source.

Finally, we derive the corresponding detector response and discuss the observational signatures of these correlated polarization patterns for ground-based interferometers and pulsar-timing arrays. The resulting structure provides a potential observational test capable of distinguishing ER from the polarization predictions of General Relativity. The structure derived here predicts a specific correlation pattern among the E(2) polarization components, which may be testable with current or future multi-detector gravitational-wave observations.
\section{Deviation tensor of gravitational waves from a binary in Extended Relativity}
\label{sec:ER-binary-retarded}

Extended Relativity (ER) describes gravitation through a \emph{deviation tensor}
$h_{\mu\nu}$ defined on a flat Minkowski background
\[
g_{\mu\nu}=\eta_{\mu\nu}-h_{\mu\nu},
\]
where $\eta_{\mu\nu}=\mathrm{diag}(1,-1,-1,-1).$ For a single source field, $h_{\mu\nu}$ is constructed covariantly from retarded source data.
In this section, we derive the ER deviation tensor $h_{\mu\nu}$ produced by a compact binary
in circular orbit, keeping all contributions up to
$O(\beta^2)$, where $\beta=v/c\ll 1$ is the orbital velocity parameter.

\subsection{ER deviation tensor for a single source}

For a point source of mass $m$ (with associated length parameter $M = Gm/c^2$), the deviation tensor
\begin{equation}\label{eq:h-def}
h_{\mu\nu}(x) = 2M\, l_\mu(x) l_\nu(x)
\end{equation}
at the spacetime observation point $x = (ct,\mathbf{x})$ is proportional to the source mass and has rank one.

The Lorentz-covariant vector $l^\mu(x)$ is constructed from retarded source data as follows. We parametrize the worldline of the source by the coordinate time $t$ of the
background Minkowski frame,
\[
x_s^\mu(t) = (ct,\mathbf{x}_s(t)).
\]
For a given spacetime point $x^\mu = (ct,\mathbf{x})$, the retarded time
$t_r = t_r(x)$ is defined by the null condition
\[
\bigl(x^\mu - x_s^\mu(t_r)\bigr)
\bigl(x_\mu - x_{s\mu}(t_r)\bigr) = 0.
\]
\[ 
t - t_r = \frac{|\mathbf{x} - \mathbf{x}_s(t_r)|}{c}.
\]

We define the \emph{ retarded position four-vector}
\begin{equation}
r^\mu(x) = x^\mu - x_s^\mu(t_r(x)),
\end{equation}
which satisfies
\[
r^\mu r_\mu = 0.
\]
The null four-vector $r$ can be written in a $1+3$ representation as 
\begin{equation}\label{eq:r-null-c1}
r^\mu=\rho(1,\hat{r}),
\end{equation} 
where $\rho=r^0$ and $\hat{r}$ has $3D$ norm one. 
The direction of propagation is given by the null vector $r^\mu$, whose
spatial part defines the unit vector $\hat{r}$.
 The  \emph{retarded four-velocity} $w^\mu$ is defined by
\begin{equation}\label{eq:w-gamma}
w^\mu=\gamma(1,\boldsymbol{\beta}),\qquad \boldsymbol{\beta} = \frac{d \mathbf{x}_s(t)}{dt}\bigg|_{t=t_r},\qquad \gamma=\frac{1}{\sqrt{1-\beta^2}}.
    \end{equation}

Then, as in \cite{FSuper}, for a single-source gravitational field, we have
\[
l^\mu
=2\alpha^{1/2}w^\mu-\alpha^{3/2}r^\mu,
\qquad
\alpha=(r\!\cdot\! w)^{-1}
\label{eq:l-def}
\]
 and from \eqref{eq:h-def}
\begin{equation}
h_{\mu\nu}
=2M\Big(
4\alpha\,w_\mu w_\nu
-2\alpha^2( w_{\mu}r_{\nu}+w_{\nu}r_{\mu})
+\alpha^3 r_\mu r_\nu
\Big).
\label{eq:h-expanded}
\end{equation}
This deviation tensor for a static point source was introduced in \cite{FStav}, where it was shown that the corresponding metric satisfies Einstein's field equations and therefore reproduces all classical tests of General Relativity. In \cite{ana}, it was demonstrated directly that this metric yields the correct predictions for particle motion, including in strong-field regimes. The extension of the deviation tensor to a non-static point source was introduced in \cite{FSuper}.

To expand the deviation tensor into powers of $\beta$, define 
\begin{equation}\label{sDef}
s:=\hat{r}\!\cdot\!\boldsymbol{\beta}=\beta (\hat{r}\!\cdot\!\hat{\beta}),
\end{equation}
which is of order $\beta$. Using \eqref{eq:r-null-c1},  \eqref{eq:w-gamma} and $\gamma=1+\tfrac12\beta^2+O(\beta^4)$,
\begin{equation}
r\!\cdot\! w =\rho\,\gamma(1-s)
\quad\Rightarrow\quad
\alpha = \frac{1}{\rho}\,\frac{1}{\gamma(1-s)}
=
\frac{1}{\rho}\Bigl(1+s+s^2-\tfrac12\beta^2\Bigr)+O(\beta^3),
\label{eq:alpha-expand-c1}
\end{equation}
where mixed terms such as $\beta^2 s=O(\beta^3)$ are dropped.
Expand\[
(1-s)^{-k}=1+k s+\tfrac12 k(k+1)s^2+O(\beta^3).
\]

Inserting these expressions into \eqref{eq:h-expanded}, one finds the far-zone expansion of the components of the deviation tensor $\tilde{h}_{\mu\nu}$ of a single source to order $O(\beta^2)$  in the lab frame basis:
\begin{equation}\label{h00}
    \tilde{h}_{00}=\frac{2M}{\rho}\Bigl(1-s-2s^2+\frac{5}{2}\beta^2\Bigr)+O(\beta^3),
\end{equation}
\begin{equation}\label{h0i}
    \tilde{h}_{0i}
=\frac{2M}{\rho}\Bigl((1+s+\frac{1}{2}\beta^2)\hat{r}_i-2\beta_{i}\Bigr)+O(\beta^3),
\end{equation}
and \begin{equation}\label{hij}
  \tilde{h}_{ij}
=\frac{2M}{\rho}\Bigl(
(1+3s+6s^2-\frac{3}{2}\beta^2)\hat{r}_i\hat{r}_k
-2(1+2s)(\beta_{i}\hat{r}_k+\beta_{j}\hat{r}_i)
+4\beta_{i}\beta_{j}
\Bigr)+O(\beta^3),  
\end{equation}
for $i,j=\{1,2,3\}$.

\subsection{Binary sum and COM condition}

Consider a binary of two stars $A$ and $B$ of masses $M_A$ and $M_B$  in circular motion. We describe the motion of the binary system in a source-adapted frame
\[
B_{\mathrm{bin}} = (\mathbf{e}_1,\mathbf{e}_2,\mathbf{e}_3),
\]
defined on a Minkowski background, with origin at the center of mass of
the system. The spatial basis is chosen such that $\mathbf{e}_3 \parallel \hat{L},$
where $\hat{L}$ is the direction of the orbital angular momentum of the binary.
Spacetime coordinates in this frame are denoted by
\[
x^\mu = (ct,\mathbf{x}) = (ct,x^1,x^2,x^3).
\]
In this frame, the motion of the two bodies is confined to the plane
orthogonal to $\hat{L}$, i.e. the $(x^1,x^2)$ plane. 

Denoting the angular velocity by $\Omega$, the positions of the stars, the sources of the gravitational field, are
\[\mathbf{x}_A(t)=R_A(\cos\Omega (t+t_0),\sin \Omega (t+t_0),0)=R_A\hat{x},\;\;\mathbf{x}_B(t)=- R_B\hat{x}, \]
where $\Omega t_0$ is the initial phase. Form COM definition 
\[R_A=\frac{M_B R}{M},\;\;R_B=\frac{M_A R}{M} ,\]
where $R=R_A+R_B$ and $M=M_A+M_B$. The velocities of the stars are the following:
\[\boldsymbol{\beta}_A(t)=\frac{R_A\Omega}{c} (-\sin \Omega (t+t_0),\cos\Omega (t+t_0),0)=\beta_A \hat{\beta}(t),\;\boldsymbol{\beta}_B(t)=\beta_B(- \hat{\beta}(t)) .\] 

We assume that the deviation tensor $h_{\mu\nu}$ of the binary is the sum of the deviation tensors of each star. To use the above formulas for the components of the deviation tensor, we need formulas for combinations of multiples of the mass of the star with $s, s^2$ and $\beta^2$. From the above formulae, we have
\[M_A\beta_A=M_A\frac{M_B R}{M}\frac{\Omega}{c}=\mu\frac{R\Omega}{c}=\mu\beta=M_B\beta_B\,,\] where the \emph{reduced mass} $\mu=\frac{M_A M_B}{M}$ and $\beta=\frac{R\Omega}{c}=\beta_A+\beta_B$. This implies that
\begin{equation}\label{bin_beta}
    M_A\boldsymbol{\beta}_A(t)+M_B\boldsymbol{\beta}_B(t)=0
\end{equation}
and
\begin{equation}\label{bin_beta2}
    M_A \beta_A^2+M_B\beta_B^2=\mu \beta^2.
\end{equation}

For the deviation tensor, we have to use the values of $w$ and $\boldsymbol{\beta}_k$ at retarded times $t_{rk}$ for each star $k$, which are:
\begin{equation}
t_{rk}
=t-\frac{\rho}{c}+\frac{\hat{r}\!\cdot\!\mathbf{x}_k(t_c)}{c}
= t_c+\delta t_k,
\qquad
t_c:=t-\frac{\rho}{c},
\qquad
\delta t_k=\frac{\hat{r}\!\cdot\!\mathbf{x}_k(t_c)}{c},
\label{eq:retarded-time}
\end{equation}
where $t_c$ is the retarded time at the center of the binary.
Hence, the source velocities at the retarded times can be expanded as
\begin{equation}
\boldsymbol{\beta}_k(t_{rk})
=\boldsymbol{\beta}_k(t_c)
+\delta t_k\,\dot{\boldsymbol{\beta}}_k(t_c)
+O(\beta^3).
\label{eq:beta-retarded}
\end{equation}
Since $\dot{\hat{\beta}}=-\Omega \hat{x}$, $\delta t_k\,\dot{\boldsymbol{\beta}}_k=-\beta_k^2(\hat{r}\cdot \hat{x})\hat{x}$, using \eqref{bin_beta} and \eqref{bin_beta2}:
\begin{equation}
M_A\,\boldsymbol{\beta}_A(t_{rA})+M_B\,\boldsymbol{\beta}_B(t_{rB})
=
M_A\,\delta t_A\,\dot{\boldsymbol{\beta}}_A(t_c)+M_B\,\delta t_B\,\dot{\boldsymbol{\beta}}_B(t_c)=-\mu\beta^2(\hat{r}\cdot \hat{x})\hat{x}.
\label{eq:retard-survival}
\end{equation}
 Thus, retardation for terms of order $\beta$ for a binary produces contributions $O(\beta^2)$. Since retardation increases the order of $\beta$, for terms of order 2 in $\beta$, we can ignore retardation. 

\subsection{Wave-frame basis}

For the analysis of gravitational-wave propagation and detector response,
it is convenient to introduce a second frame, the wave frame
\[
B_{\mathrm{w}} = (\mathbf{e}'_1,\mathbf{e}'_2,\mathbf{e}'_3),
\]
in which the spatial basis is adapted to the direction of propagation
of the wave.

We define
\[
\mathbf{e}'_3 = \hat{r},
\]
where $\hat{r}$ is the unit vector in the direction of propagation of the
wave at the observation point. The remaining basis vectors are constructed
relative to the direction of the angular momentum $\hat{L}$.
The \emph{inclination angle} $\theta$ is defined by
\begin{equation}
\cos\theta = \hat{r}\cdot \hat{L}.
\end{equation}

The vector
\[
\frac{\hat{r}\times \hat{L}}{|\hat{r}\times \hat{L}|}
\]
lies in the orbital plane. By an appropriate rotation within this plane,
we choose
\[
\mathbf{e}'_2 = (0,1,0), \qquad
\mathbf{e}'_1 = \mathbf{e}'_2 \times \mathbf{e}'_3.
\]
If $\hat{r}\times \hat{L}=0$, we chose $\mathbf{e}'_2 =\mathbf{e}_2$.

With this choice, the spatial basis vectors in the wave frame take the form
\begin{equation}\label{Bw_new}
\mathbf{e}'_3 = (\sin\theta, 0, \cos\theta), \qquad
\mathbf{e}'_1 = (\cos\theta, 0, -\sin\theta), \qquad
\mathbf{e}'_2 = (0,1,0).
\end{equation}
Spacetime coordinates in the wave frame are denoted by
\[
x'^\mu = (ct,\mathbf{x}') = (ct,x'^1,x'^2,x'^3).
\]
This frame provides a natural basis for expressing the deviation tensor
in the wave zone and for analyzing its polarization structure.

Define the retarded phase by
\begin{equation}\label{phaseRet}
 \psi=\Omega (t_c+t_0)=\Omega(t-\rho/c +t_0),
\qquad
.   
\end{equation}
then
\[\hat{x}
=(\cos\psi,\sin\psi,0)\quad\hbox{ and }\quad\hat{\beta}
=(-\sin\psi,\cos\psi,0),\]
implying that
\[\hat{r}\cdot\hat{x}=\sin \theta\cos\psi\quad\hbox{ and }\quad \hat{r}\cdot\hat{\beta} =-\sin \theta \sin \psi\,.\]
Using these and \eqref{eq:retard-survival}, we have
\[ M_A\,s_A(t_{rA})+M_B\,s_B(t_{rB})=\hat{r}\cdot (M_A\,\boldsymbol{\beta}_A(t_{rA})+M_B\,\boldsymbol{\beta}_B(t_{rB}))\]
\begin{equation}\label{sumRetar}
=-\mu \beta^2(\hat{r}\cdot\hat{x})^2=-\mu \beta^2\sin^2\theta\cos^2\psi=-\mu \beta^2\frac{\sin^2\theta}{2}(1+\cos2\psi).\end{equation}

Since retardation increases the order of $\beta$, for terms of order 2 in $\beta$, we can ignore retardation. From \eqref{sDef} and \eqref{bin_beta2}, it follows that to order $O(\beta^2)$, we have
\begin{equation}\label{sum2}
M_A\,s_A^2(t_{rA})+M_B\,s_B^2(t_{rB})=M_A\,s_A^2(t_c)+M_B\,s_B^2(t_c)=\mu \beta^2\sin^2\theta\sin^2\psi=\mu \beta^2\frac{\sin^2\theta}{2}(1-\cos2\psi).  
\end{equation}

\subsection{Components of binary deviation tensor in wave-frame basis }

We denote by $h_{\mu\nu}(x)$ the deviation tensor of the binary in the wave-frame basis $B_{\mathrm{w}}$, evaluated at the spacetime point $x = (ct,\mathbf{x})$. It is given by the sum of the deviation tensors associated with the two stars, labeled by $k \in \{A,B\}$.

Using \eqref{h00}, \eqref{sumRetar}, \eqref{sum2} and \eqref{bin_beta2}
\[h_{00}=\sum_k\tilde{h}_{00}^k=\sum_k\frac{2M_k}{\rho}(1-s_k-2s_k^2+\frac{5}{2}\beta^2_k)=\frac{2M}{\rho}
+\frac{\mu\beta^2}{\rho}\bigl(5+\sin^2\theta(-1+3\cos2\psi)\bigr).\]
Using \eqref{h0i}, \eqref{sDef}, \eqref{sumRetar} and \eqref{bin_beta2}
\[h_{03}=\sum_k\tilde{h}^k_{0i}\hat{r}^i=\sum_k\frac{2M_k}{\rho}(1-s_k+\frac{1}{2}\beta^2)
=\frac{2M}{\rho}
+\frac{\mu\beta^2}{\rho}\bigl(1+\sin^2\theta(1+\cos2\psi)\bigr).\]
Using \eqref{h0i} and \eqref{eq:retard-survival}
\[h_{01}=\sum_k\tilde{h}^k_{0i}\mathbf{e}'^i_1=\sum_k\frac{2M_k}{\rho}(-2\boldsymbol{\beta}_k\cdot \mathbf{e}'_1)=\frac{-4\mu\beta^2}{\rho}(\hat{r}\cdot \hat{x})(\hat{x}\cdot \mathbf{e}'_1)
=\frac{2\mu\beta^2}{\rho}\sin\theta\cos\theta\bigl(1+\cos2\psi\bigr).\]
Using \eqref{h0i} and \eqref{eq:retard-survival}
\[h_{02}=\sum_k\tilde{h}^k_{0i}\mathbf{e}'^i_2=\sum_k\frac{2M_k}{\rho}(-2\boldsymbol{\beta}_k\cdot \mathbf{e}'_2)
=\frac{2\mu\beta^2}{\rho}\sin\theta\sin2\psi\]
Using \eqref{hij}, \eqref{sDef},\eqref{sumRetar}, \eqref{sum2} and \eqref{bin_beta2} 
\[h_{33}=\sum_k\tilde{h}^k_{ij}\hat{r}^i\hat{r}^j=\sum_k\frac{2M_k}{\rho}(1-s_k+2s_k^2-\frac{3}{2}\beta^2_k)
=\frac{2M}{\rho}
+\frac{\mu\beta^2}{\rho}\bigl(-3+\sin^2\theta(3-\cos2\psi)\bigr).\]
Similarly,
\[h_{13}
=\sum_k\tilde{h}^k_{ij}\hat{r}^i\mathbf{e}'^j_1=\sum_k\frac{2M_k}{\rho}(-2\boldsymbol{\beta}_k\cdot \mathbf{e}'_1)=h_{01}\]
\[h_{23}=\sum_k\tilde{h}^k_{ij}\hat{r}^i\mathbf{e}'^j_2=\sum_k\frac{2M_k}{\rho}(-2\boldsymbol{\beta}_k\cdot \mathbf{e}'_2)=h_{02}
.\]
Since for second order terms we can ignore the retardation,
using \eqref{bin_beta2}:
\[h_{11}=\sum_k\tilde{h}^k_{ij}\mathbf{e}'^i_1\mathbf{e}'^j_1=\sum_k\frac{2M_k}{\rho}4\beta_k^2(\hat{\beta}\cdot \mathbf{e}'_1)^2
=\frac{4\mu\beta^2}{\rho}\cos^2\theta\bigl(1-\cos2\psi\bigr),\]
\[h_{12}=\sum_k\tilde{h}^k_{ij}\mathbf{e}'^i_1\mathbf{e}'^j_2
=\sum_k\frac{2M_k}{\rho}4\beta_k^2(\hat{\beta}\cdot \mathbf{e}'_1)(\hat{\beta}\cdot \mathbf{e}'_2)
=-\frac{4\mu\beta^2}{\rho}\cos\theta\,\sin2\psi.\]

\[h_{22}=\sum_k\tilde{h}^k_{ij}\mathbf{e}'^i_2\mathbf{e}'^j_2 = \sum_k\frac{2M_k}{\rho}4\beta_k^2(\hat{\beta}\cdot \mathbf{e}'_2)^2=\frac{4\mu\beta^2}{\rho}\bigl(1+\cos2\psi\bigr)\]

Finally, the deviation tensor of a binary to $O(\beta^2)$ is
\begin{equation}\label{DevFin}
h_{\mu\nu}=\frac{2M}{\rho}h_{\mu\nu}^{(0)}+\frac{\mu\beta^2}{\rho}\big(h_{\mu\nu}^{(1)}(\theta)+C_{\mu\nu}(\theta)\cos2\psi+S_{\mu\nu}(\theta)\sin 2\psi \big),
\end{equation}
where
\[h_{\mu\nu}^{(0)}=
\begin{pmatrix}
1 & 0 & 0 & 1\\
0 & 0 & 0 & 0\\
0 & 0 & 0 &0\\
1 & 0& 0 & 1
\end{pmatrix} ,\;\;h_{\mu\nu}^{(1)}(\theta)=
\begin{pmatrix}
4+\cos^2\theta & \sin2\theta & 0 & 2-\cos^2\theta\\
\sin2\theta & 4\cos^2\theta & 0 & \sin2\theta\\
0 & 0 & 4 & 0\\
2-\cos^2\theta &\sin2\theta &0 & -3\cos^2\theta
\end{pmatrix}\] 
and
\[C_{\mu\nu}(\theta)=
\begin{pmatrix}
3\sin^2\theta & \sin2\theta &0 &\sin^2\theta\\
\sin2\theta & -4\cos^2\theta& 0 & \sin2\theta\\
0 & 0 & 4 &0\\
\sin^2\theta&\sin2\theta &0& -\sin^2\theta
\end{pmatrix},\]
\begin{equation}\label{CSten}
S_{\mu\nu}(\theta)=
\begin{pmatrix}
0 & 0 &2\sin\theta  &0\\
0 & 0 & -4\cos\theta  &0\\
2\sin\theta & -4\cos\theta & 0 &2\sin\theta \\
0 &0 &2\sin\theta & 0
\end{pmatrix}
\end{equation}

We decomposed the tensor into 4 components. All terms proportional to $\frac{1}{\rho}$. The first one is the only one of order 0 in beta. It is equal to the deviation tensor of a total mass $M$ positioned at the center of mass of the binary. All other terms are of order $O(\beta^2)$ and connected to the reduced mass $\mu$. The second terms are static in time, while the last two oscillate with frequency $2\Omega$. The tensor part of the last 3 terms depend only on $\theta$ and remain constant in the far region. Note that the relativistic trace  $\eta^{\mu\nu} h_{\mu\nu}$ of the deviation tensor and of each component  vanishes.

Our derivation was based on the single-source deviation tensor \eqref{eq:h-expanded}, which is true for the gravitational field of a black hole. For fields of spherically symmetric bodies, such as neutron stars, it was shown in \cite{FK} that this formula gives a good approximation, and a finite-size correction was identified. Thus, our formula for the binary deviation tensor is also valid for neutron star binaries, except during the final merger, when the finite-size correction becomes significant.


\section{Tidal matrix and wave polarization}
\label{sec:ER-tidal-retarded}

A gravitational-wave polarization describes the pattern of relative acceleration induced by spacetime curvature in a small cloud of freely falling test particles. This physical effect is encoded in a gauge-invariant and observer-independent way by the tidal matrix.

The tidal matrix (also called the electric component of the Riemann tensor) is defined as
\[
\mathcal{E}_{ab}:=R_{0a0b},\qquad a,b\in\{1,2,3\},
\]
where $R$ is the Riemann tensor. We calculate this matrix for the binary field defined by the deviation tensor \eqref{DevFin}.
The components of the tidal matrix can be directly mapped onto the standard E(2) polarization basis used in gravitational-wave data analysis, allowing a direct comparison between the ER predictions and the six polarization patterns commonly considered in detector response studies.
Since the ER metric is $g_{\mu\nu}=\eta_{\mu\nu}-h_{\mu\nu},$ this matrix at any point $x$ is
\begin{equation}
R_{0a0b}
=\frac12\Big(
-\partial_0\partial_a h_{0b}
-\partial_0\partial_b h_{0a}
+\partial_0^2 h_{ab}
+\partial_a\partial_b h_{00}
\Big).
\label{eq:R0a0b-general}
\end{equation}

In the radiation zone, using  the wave-frame basis  $B_{\mathrm{w}}$, under the leading plane-wave approximation, angular derivatives $\partial/\partial x'^1$ and $\partial/\partial x'^2$, as well as derivatives of $1/\rho$, may be neglected. In this basis, with $\mathbf{e}'_3 = \hat{r}$, the observation point lies along the propagation direction, $\mathbf{x} = \rho\,\hat{r}$, implying
\[
x'^1 = x'^2 = 0, \qquad x'^3 = \rho.
\]
In this regime, the deviation tensor defined in \eqref{DevFin} depends predominantly on the retarded phase $\psi$, introduced in \eqref{phaseRet}, which can be written as
\[
\psi = \frac{\Omega}{c}(ct - \rho +ct_0) = \frac{\Omega}{c}(x'^0 - x'^3+ct_0).
\]
Consequently, to leading order,
\[
\partial_3 = -\partial_0, 
\qquad 
\partial_1 = \partial_2 = 0.
\]

Under this approximation, the tidal components simplify as follows:
\begin{itemize}
\item For transverse indices $i,j\in\{1,2\}$,
\begin{equation}
\mathcal{E}_{ij}= \tfrac12\,\partial_0^2 h_{ij}.
\label{eq:EAB}
\end{equation}
\item For mixed components,
\begin{equation}
\mathcal{E}_{3i}
= \tfrac12\big(
-\partial_0\partial_3 h_{0i}
+\partial_0^2 h_{3i}
\big)
= \partial_0^2 h_{0i}\,,
\label{eq:ErA}
\end{equation}
where the last equality uses $h_{3i}=h_{0i}$ at this order.

\item For the radial component,
\begin{equation}
\mathcal{E}_{33}
\simeq \tfrac12\,\partial_0^2\big(
2h_{03}+h_{33}+h_{00}
\big).
\label{eq:Err}
\end{equation}
\end{itemize}

The relevant second derivatives are as follows:
\[
\partial_0^2(\cos2\psi)=-(2\Omega/c)^2\cos2\psi,
\qquad
\partial_0^2(\sin2\psi)=-(2\Omega/c)^2\sin2\psi .
\]
Define the overall amplitude
\begin{equation}\label{Ampl}
\mathcal{A}:=\frac{8\mu\beta^2\Omega^2}{\rho c^2}.
\end{equation}
Using \eqref{eq:EAB}--\eqref{eq:Err}, we obtain the  tidal matrix
\begin{equation}\label{eq:tidal-matrix-final}
\mathcal{E}_{ab}=
\mathcal{A}
\begin{pmatrix}
\cos^2\theta\,\cos2\psi
&
\cos\theta\,\sin2\psi
&
\sin\theta\cos\theta\,\cos2\psi
\\[4pt]
\cos\theta\,\sin2\psi
&
-\cos2\psi
&
-\sin\theta\,\sin2\psi
\\[4pt]
\sin\theta\cos\theta\,\cos2\psi
&
-\sin\theta\,\sin2\psi&
 -\sin^2\theta\,\cos2\psi
\end{pmatrix}.    
\end{equation}

Equation~\eqref{eq:tidal-matrix-final} shows that
the curvature contains not only the transverse tensor components, but also
non-vanishing radial and mixed components.
These terms correspond to longitudinal and vector-like tidal components of the curvature tensor at order $O(\beta^2)$ and are absent if the retarded times of the two sources are
artificially identified.
This completes the derivation of the physical tidal matrix associated with
gravitational radiation from a circular binary in Extended Relativity.

Using the tidal matrix, and that $\theta$ is constant in radiation region and $\psi$ depends on time, we obtain the following polarization modes
\begin{equation}
\begin{aligned}
h_+(t)
&=\tfrac12\big(\mathcal{E}_{11}-\mathcal{E}_{22}\big)
=\frac{\mathcal{A}}{2}\bigl(1+\cos^2\theta\bigr)\cos2\psi,\\[4pt]
h_\times(t)
&=\mathcal{E}_{12}
=\mathcal{A}\cos\theta\,\sin2\psi,\\[4pt]
h_b(t)
&=\tfrac12\big(\mathcal{E}_{11}+\mathcal{E}_{22}\big)
=-\frac{\mathcal{A}}{2}\sin^2\theta\,\cos2\psi,\\[4pt]
h_x(t)
&=\mathcal{E}_{13}
=\mathcal{A}\sin\theta\cos\theta\,\cos2\psi,\\[4pt]
h_y(t)
&=\mathcal{E}_{23}
=-\mathcal{A}\sin\theta\,\sin2\psi,\\[4pt]
h_\ell(t)
&=\mathcal{E}_{33}
=-\mathcal{A}\sin^2\theta\,\cos2\psi.
\end{aligned}
\label{eq:ER-E2-amplitudes}
\end{equation}
Each polarization mode $h_P$ has a polarization amplitude $A_P(\theta)$, which is constant in the radiation region,  multiplied by $\cos(2\psi)$ or $\sin(2\psi)$. 
These expressions show that the ER radiation field corresponds to a constrained mixture of the six E(2) polarization patterns, where the amplitudes are not independent but are fixed by the binary inclination and source dynamics.

\begin{figure}[t]
  \centering
  \includegraphics[width=0.9\linewidth]{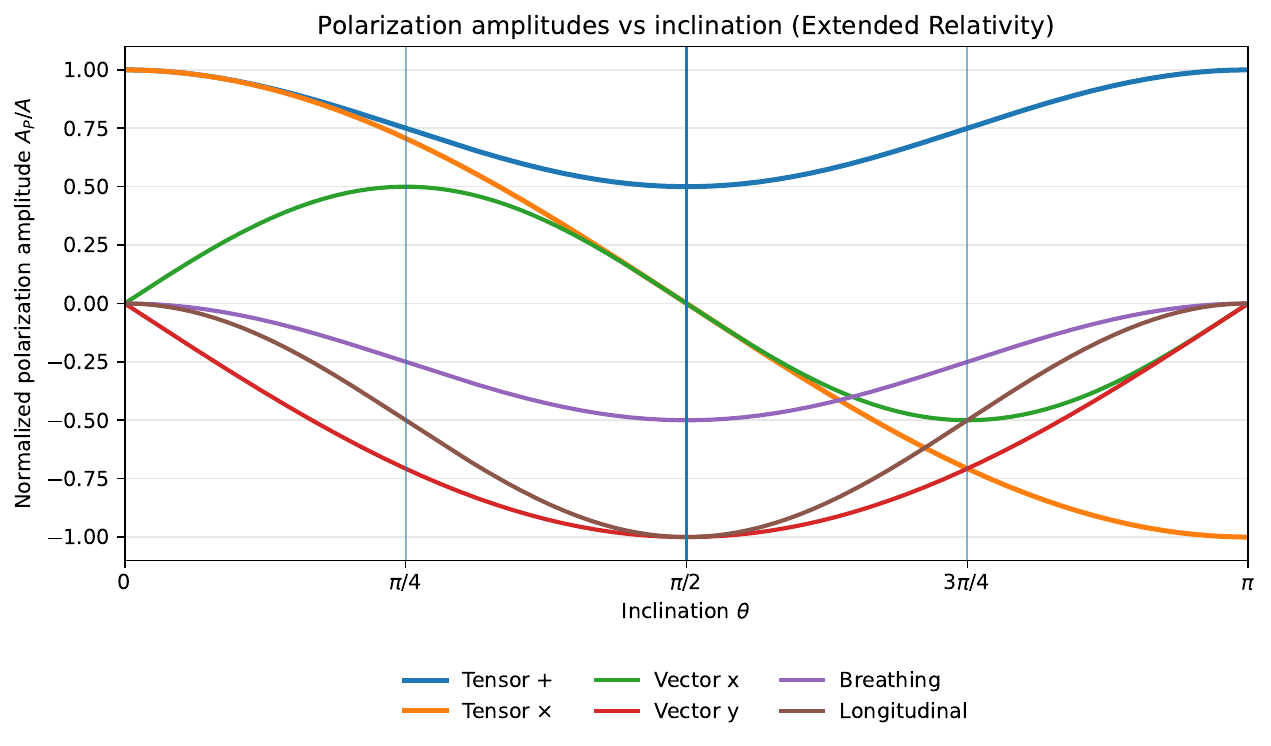}
  \caption{Angular dependence of the polarization amplitudes predicted by
Extended Relativity for a circular binary, normalized by the overall
amplitude A defined in Eq. \eqref{Ampl}. The tensor components $(+,\times)$ reproduce
the standard quadrupole structure of General Relativity, while ER
predicts additional breathing, vector, and longitudinal tidal
components whose amplitudes are fixed functions of the inclination
angle $\theta$. $\theta=0$— face-on binary, $\theta=\pi/2$—edge-on binary, $\theta =\pi$-face-on (opposite orientation) }
  \label{XX}
\end{figure}

The inclination dependence of the polarization amplitudes is illustrated in  Fig.\ref{XX}, showing that ER predicts a correlated mixture of tensor, vector, and scalar tidal components whose relative strengths are fixed by the binary orientation.
 ER predicts that non-tensor tidal components vanish for face-on binaries but become significant for inclined systems. 

Equation~\eqref{eq:ER-E2-amplitudes} has the following interpretation.

\begin{itemize}
\item \textbf{Tensor sector.}
The components $A_+$ and $A_\times$ describe the familiar transverse shear.
Their angular dependence $(1+\cos^2\theta)/2$ and $\cos\theta$ matches the standard quadrupole pattern for circular binaries, showing that ER reproduces GR-like tensor waveforms in this sector. For face-on binaries ($\theta=0,\pi$)  all non-tensor polarization vanishes, and the ER polarization agrees with the GR one.

\item \textbf{Scalar breathing.}
The nonzero $A_b$ describes an isotropic expansion/contraction in the plane transverse
to propagation. It is maximal for
edge-on binaries ($\theta=\pi/2$). This mode corresponds to the non-vanishing trace
in the transverse tidal block.

\item \textbf{Vector-like couplings.}
The amplitudes $A_x$ and $A_y$ represent tidal couplings involving the propagation
direction. In the present derivation, they arise from the \emph{retardation of
the linear-in-$\beta$ terms} in the binary field. These terms vanish if the two retarded
times are artificially identified, hence they encode a genuinely retarded two-body
effect in ER.

\item \textbf{Longitudinal scalar.}
The component $A_\ell=\mathcal{E}_{33}$ describes tidal stretching/compression along the
propagation direction. In ER it is nonzero here, indicating the presence of a true
longitudinal curvature component at $O(\beta^2)$.

\item \textbf{Physical significance.}
Because $\mathcal{E}_{ab}$ controls the relative acceleration, the amplitudes \eqref{eq:ER-E2-amplitudes} are directly measurable
in principle by suitably configured detector networks. In particular, ER predicts additional radiative channels beyond $(+,\times)$, which implies that future polarization
measurements (breathing, vector, and longitudinal responses) provide  a clear way to distinguish
 ER from GR, even when the tensor modes coincide.

\item \textbf{ER polarization from edge-on binaries}
The polarization of GW from edge-on binaries ($\theta=\pi/2$) predicted by ER is the following:
\[h_+=\frac{\mathcal{A}}{2}\cos2\psi,\; h_\times=0,\; h_b=-\frac{\mathcal{A}}{2}\cos2\psi\]
\[h_x=0,\; h_y=-\mathcal{A}\sin2\psi,\;h_\ell=-\mathcal{A}\cos2\psi,\]
which is very different from the GR polarization.
\end{itemize}

\section{ Detector response to  GW }

The polarization content described above fits naturally within the
morphology-independent framework of Chatziioannou, Yunes and Cornish \cite{mixed}, which decomposes the detector
response into the six E(2) polarization modes with arbitrary,
a~priori~independent time-dependent amplitudes.
In this language, the ER model corresponds to a restricted subclass of
mixed-polarization signals involving vector-$x$, vector-$y$, and
scalar--longitudinal modes.
Unlike the generic parameterization adopted in
morphology-independent searches, the ER model predicts fixed phase
relations and inclination-dependent amplitude ratios among these modes,
with all components oscillating at twice the orbital frequency.
As a result, existing analyzes that marginalize over independent
polarization amplitudes do not directly constrain this specific,
phase-locked vector--longitudinal signal, motivating a targeted
polarization analysis tailored to the ER predictions.


A ground-based laser interferometric detector measures gravitational
waves through the differential change in the optical path lengths of
its two orthogonal arms.
For a plane gravitational wave, the measured strain at the detector $d$ may be written as
\begin{equation}\label{strainGen}
h^d(t)
= \sum_{P} F_P^d(B)\, A_P(t),
\end{equation}
where $P \in \{+, \times, x, y, b, l\}$ labels the six possible E(2)
polarization modes, the antenna pattern function $F_P^d(B)$ depends on the antenna configuration, polarization, and wave basis $B_{\mathrm{w}}$, and $A_P(t)$ are the corresponding polarization
amplitudes.

To define the antenna pattern function, we use a fixed lab frame coordinate system (mainly ECEF) in which the antenna pattern function is:
\begin{equation}
F_P^d(\hat{\mathbf{r}})
= D^d_{ab} \, e^{ab}_P(\hat{\mathbf{r}}),
\end{equation}
where $D^d_{ab}$ is called the detector tensor and $e^{ab}_P$ are the E(2) polarization tensors. The detector tensor is defined from the unit vectors
$\hat{\mathbf{u}}^d=(\hat{u}^d_1,\hat{u}^d_2 ,\hat{u}^d_3)$ and $\hat{\mathbf{v}}^d=(\hat{v}^d_1,\hat{v}^d_2 ,\hat{v}^d_3)$ along the two arms of the interferometer in the lab frame as:
\begin{equation}
D^d_{ab}
= \frac{1}{2}
\left( \hat{u}^d_a \hat{u}^d_b - \hat{v}^d_a \hat{v}^d_b \right)
\end{equation}
 Denoting the lab frame coordinates of the wave basis $B_{\mathrm{w}}$ as $\mathbf{e}'_k=(e_k^1 ,e_k^2, e_k^3)$, the six E(2) polarization basis tensors are
\begin{align}
e_{+}^{ab} &= e_1^a e_1^b-e_2^a e_2^b, &
e_{\times}^{ab} &= e_1^ae_2^b+e_2^ae_1^b,
\nonumber\\
e_{x}^{ab} &= e_3^ae_1^b+e_1^ae_3^b, &
e_{y}^{ab} &= e_3^ae_2^b+e_2^ae_3^b,
\nonumber\\
e_{b}^{ab} &= e_1^ae_1^b+e_2^ae_2^b, &
e_{\ell}^{ab} &= e_3^ae_3^b.
\label{eq:E2-basis}
\end{align}
They correspond respectively to tensor $(+,\times)$, vector $(x,y)$, scalar breathing $(b)$,
and scalar longitudinal $(\ell)$ polarizations.

By use of \eqref{eq:ER-E2-amplitudes}, the strain \eqref{strainGen} measured by detector $d$ can be decomposed as
\begin{equation}
h^d(t)
= A_c^d(\theta)\,\cos(2\Omega(t-\rho/c)) +A_s^d(\theta)\,\sin(2\Omega(t-\rho/c)),
\end{equation}
where
\begin{equation}\label{Ac}
  A_c^d(\theta)=\mathcal{A}/2(F_+^d(\hat{\mathbf{r}})(1+\cos^2\theta)-F_b^d(\hat{\mathbf{r}})\sin^2\theta+F_x^d(\hat{\mathbf{r}})2\sin \theta-F_\ell^d(\hat{\mathbf{r}})2\sin^2\theta)  
\end{equation} and
\begin{equation}\label{As}
  A_s^d(\theta)=\mathcal{A}(F_\times^d(\hat{\mathbf{r}})\cos\theta -F_y^d(\hat{\mathbf{r}})\sin\theta).  
\end{equation}
This expression can be written equivalently as
\begin{equation}\label{strainF}
h^d(t)
= A_T^d(\theta)
\cos\!(2\Omega t_r + \Phi^d(\theta)),
\end{equation}
where the detector-dependent amplitude $A_T^d$ and phase $\Phi^d$ are
given by
\begin{equation}
  (A_T^d)^2(\theta) =
\left(F_c^d (\theta)\right)^2
+ \left(F_s^d (\theta)\right)^2, \;\;\;
\tan\Phi^d (\theta)=
\frac{F_s^d(\theta) }{F_c^d(\theta)}.  
\end{equation}

For two detectors $a$ and $b$ separated by the vector $\mathbf{r}_{ab}$, the retarded times satisfy
\begin{equation}
t_c^b = t_c^a + \frac{\mathbf{r}_{ab}\cdot\hat{\mathbf{r}}}{c},
\end{equation}
which gives rise to a geometrical time delay between the detectors.

The observed strain signals therefore differ only by:
\begin{enumerate}
\item the antenna pattern functions $F_P^d$,
\item the detector-dependent amplitudes $A_T^d$ and phase shifts $\Phi^d$,
\item the geometrical time delays between detectors.
\end{enumerate}

In summary, the detector response in the ER framework retains the standard linear decomposition in terms of E(2) polarization modes, but the corresponding amplitudes are not independent. Instead, they are phase-locked and determined by the binary inclination angle and source dynamics, leading to a constrained and predictive signal structure. As a result, the observed strain in different detectors is fully characterized by a reduced set of parameters, with differences arising only from the antenna pattern functions and geometrical time delays. This distinguishes ER from morphology-independent approaches, where polarization amplitudes are treated as arbitrary functions of time. Consequently, a dedicated analysis that incorporates the ER constraints—rather than marginalizing over independent modes—provides a direct and testable way to probe the presence of correlated vector and longitudinal components in gravitational-wave data.

\section{Pulsar timing array response}

Pulsar timing arrays (PTAs) probe gravitational waves through their effect
on the propagation of photons along null geodesics. In contrast to
interferometric detectors, which measure the local tidal field
$E_{ab} = R_{0a0b}$, PTA observables are governed by the connection
coefficients $\Gamma^\mu_{\alpha\beta}$ and therefore depend on first
derivatives of the deviation tensor $h_{\mu\nu}$.

\subsection{Photon propagation and redshift}

Let $k^\mu$ be the null four-vector tangent to the photon trajectory,
normalized such that $k^\mu = \omega(1,-\hat{n})$,
where $\hat{n}$ is the unit vector from the Earth to the pulsar. Then, $k^\mu k_\mu = 0,$
where indices are raised and lowered using the Minkowski metric $\eta_{\mu\nu}$. The photon
propagation is described by the geodesic equation
\begin{equation}
\frac{dk^\mu}{d\lambda} + \Gamma^\mu_{\alpha\beta} k^\alpha k^\beta = 0,
\end{equation}
where $\lambda$ is an affine parameter.

The observable is the fractional frequency shift
\begin{equation}
z = \frac{\delta \nu}{\nu} = \frac{\Delta(k_\mu u^\mu)}{k_\nu u^\nu},
\end{equation}
where $u^\mu$ is the four-velocity of the observer.

\subsection{Reduction to boundary terms}

In the radiation zone, using wave-frame basis $B_{\mathrm{w}}$, the deviation tensor of the binary is given by
Eq.~\eqref{DevFin} depends on spacetime
coordinates only through the phase $\psi$.
Using
\[
\partial_\mu \psi = \frac{\Omega}{c}\,\frac{r'_\mu}{\rho},
\]
and neglecting derivatives of $1/\rho$, the derivatives
$\partial_\mu h_{\alpha\beta}$ are proportional to $r'_\mu$.
As a result, when contracted with the photon direction,
the Christoffel symbols reduce effectively to
\begin{equation}
\Gamma_{\mu\alpha\beta} k^\alpha k^\beta
\simeq \frac{1}{2}\,\partial_\mu h_{\alpha\beta}\, k^\alpha k^\beta,
\end{equation}
up to terms proportional to $k_\mu$, which do not contribute to the
observed frequency shift.

Substituting
into the geodesic equation for $k_\mu$,
\begin{equation}
\frac{d k_\mu}{d\lambda}
=
-\frac{1}{2}\partial_\mu h_{\alpha\beta}\,k^\alpha k^\beta,
\end{equation}
one finds that the right-hand side is proportional to
$\psi_{,\mu}$ and therefore becomes a total derivative along the photon
trajectory.

As a result, integration along the null ray from the pulsar ($p$) to the
Earth ($e$) yields
\begin{equation}
\Delta k_\mu
=
k_\mu(e) - k_\mu(p)
\propto
\psi_{,\mu}
\Bigl[
k^\alpha k^\beta h_{\alpha\beta}
\Bigr]_p^e.
\end{equation}

Thus, although photon propagation is governed by the Christoffel symbols,
the observable frequency shift depends only on the deviation tensor
evaluated at the endpoints of the trajectory.

\subsection{PTA response}

For an observer at rest in the wave frame, $u^\mu = (1,\mathbf{0})$, the
fractional frequency shift takes the form
\begin{equation}
z(t)
\propto
\Bigl[
k^\alpha k^\beta h_{\alpha\beta}
\Bigr]_p^e,
\end{equation}
showing that the PTA response is determined by the projection of the
deviation tensor along the photon direction.

Substituting the binary deviation tensor \eqref{DevFin}, we obtain
\begin{equation}
z(t) = C(\hat{n},\hat{r},\theta)\cos(2\psi)
+
S(\hat{n},\hat{r},\theta)\sin(2\psi),
\end{equation}
where the coefficients are determined by the contraction
$k^\mu k^\nu h_{\mu\nu}$.

In contrast to the transverse--traceless formulation of General Relativity,
this contraction involves all components of the deviation tensor,
including $h_{00}$, $h_{0a}$, and $h_{ab}$. Consequently, PTA observations
are sensitive to temporal, longitudinal, and mixed components of the
gravitational field.

The intrinsic rotation of the pulsar produces a deterministic signal that is
not correlated between different pulsars. Since PTA measurements rely on
correlated timing residuals, these rotational contributions do not affect the
correlation function and can be neglected in the present analysis.

\subsection{Angular correlations}

The PTA observable is the cross-correlation of timing residuals between
pairs of pulsars. For two pulsars separated by an angle $\zeta$, the
correlation function is defined as
\begin{equation}
\Gamma(\zeta) = \langle z_a(t)\, z_b(t) \rangle.
\end{equation}
The polarization-dependent correlation functions $\Gamma_P(\zeta)$ have been
derived and studied in the context of pulsar timing arrays in
\cite{HellingsDowns1983,Lee2008,Gair2015,NANOGrav2023,Antoniadis2023},
and provide a standard framework for testing deviations from the
transverse tensor prediction.

In the ER framework, the correlation function inherits the constrained
polarization structure derived in Section~3. It can therefore be expressed
as
\begin{equation}
\Gamma_{\mathrm{ER}}(\zeta;\theta)
=
\frac{
W_T(\theta)\,\Gamma_{\mathrm{HD}}(\zeta)
+
W_V(\theta)\,\Gamma_{V}(\zeta)
+
W_b(\theta)\,\Gamma_{b}(\zeta)
+
W_\ell(\theta)\,\Gamma_{\ell}(\zeta)
}{
W_T(\theta)+W_V(\theta)+W_b(\theta)+W_\ell(\theta)
},
\end{equation}
where the weights $W_P(\theta)$ are determined by the polarization
amplitudes in \eqref{eq:ER-E2-amplitudes}.

This formulation shows that PTA measurements provide a complementary probe
of the gravitational field, sensitive to components of the deviation tensor
that do not contribute to the transverse tidal response measured by
interferometric detectors.

\section{Summary and conclusions}

In this work, we have analyzed the polarization structure of gravitational
waves in the framework of Extended Relativity (ER), using the deviation
tensor as the fundamental observable quantity. Starting from the point-source
solution, we derived the gravitational radiation field of a compact binary
system in circular motion in the wave zone and expressed the deviation tensor \eqref{DevFin} in a form that
separates its static and oscillatory components.

A key result of this analysis is that the deviation tensor admits a decomposition
in which the spacetime dependence is carried entirely by the retarded phase,
while the tensorial coefficients depend only on the inclination angle of the
binary. This structure enables a direct and transparent computation of 
the tidal matrix \eqref{eq:tidal-matrix-final}, polarization modes \eqref{eq:ER-E2-amplitudes} and the connection coefficients, providing a unified framework
for describing the response of different classes of detectors.

For interferometric detectors, the observable signal is governed by the tidal
matrix $E_{ab} = R_{0a0b}$, which depends on second derivatives of the deviation
tensor. In contrast, pulsar timing arrays probe the propagation of photons along
null geodesics and are therefore sensitive to the connection coefficients,
i.e. to first derivatives of $h_{\mu\nu}$. We have shown that, due to the specific
phase dependence of the ER solution, integration of the geodesic equation
reduces to boundary terms, and the PTA response is determined by the projection
$k^\mu k^\nu h_{\mu\nu}$ evaluated at the emission and reception points.

An important consequence of the ER framework is that the polarization components
of the gravitational wave are not independent. Instead, the amplitudes of the
tensor, vector, and scalar contributions are fixed by the binary geometry,
leading to a constrained polarization structure, see Figure \ref{XX}. This feature distinguishes ER
from morphology-independent approaches in which polarization modes are treated
as independent degrees of freedom.

As a result, both interferometric and PTA observables can be expressed in terms
of the same underlying deviation tensor, with differences arising only from the
geometric projection relevant to each detector. This leads to a restricted family
of angular correlation functions in PTA observations, characterized by a single
physical parameter, the inclination angle.

The formulation presented here provides a consistent and unified description of
gravitational-wave observables in ER, and establishes a direct connection between
the theoretical polarization structure and observable quantities. The implications for data analysis and observational tests will be investigated in future work.

Our model predicts a mixture of transverse tensor and additional non-tensor
polarizations. As noted in Ref.~\cite{Yunes} (p.~142), the LIGO/Virgo Collaboration
found that pure scalar, pure vector, and vector–scalar mixed hypotheses are
strongly disfavored, whereas mixed hypotheses that include tensor modes cannot
be ruled out. Furthermore, as discussed in Ref.~\cite{Yunes} (p.~170) in the
context of pulsar-timing data, models exhibiting multiple spatial correlations
that include transverse--traceless modes remain theoretically viable.

\section{Acknowledgements}

The author thanks Sergei Kopeikin, Tzvi Scarr and referees for constructive comments.


\begin{thebibliography}{95}
\bibitem{Einstein16} A. Einstein, “Approximative integration of the field equations of gravitation”, Sitzungsberichte
Preußische Akademie der Wissenschaften Berlin (Math. Phys.) 688 (1916).
\bibitem{Einstein18} A. Einstein, “On gravitational waves”, Sitzungsberichte Preußische Akademie der Wissenschaften Berlin (Math. Phys.) 154 (1918).
\bibitem{Thorne} K. S. Thorne, Gravitational Radiation, in 300 Years of Gravitation, eds. Hawking and Israel (Cambridge, 1987).
\bibitem{Isi} M. Isi, M. Pitkin and A. J. Weinstein ``Probing dynamical gravity with the polarization of continuous gravitational waves" \emph{Phys. Rev. D} 96, 042001 (2017)
\bibitem{Eardley} D. M. Eardley, D. L. Lee, A. P. Lightman, R. V. Wagoner, and C. M. Will, \emph{ Phys. Rev. Lett.} 30, 884 (1973)
\bibitem{Eardley2} Eardley, D. M. and Lee, D. L. and Lightman, A. P. and Wagoner, R. V. and Will, C. M., ``
  Gravitational-wave observations as a tool for testing relativistic gravity", \emph{Phys. Rev. D.}, 8 , 3308 (1973)
\bibitem{Nishizawa} Nishizawa \emph{et al.} ``Probing non-tensorial polarizations of stochastic gravitational-wave backgrounds" \emph{Phys. Rev. D.}, 79, 082002 (2009)
 \bibitem{Abbott} B. P. Abbott \emph{et al.} ``Observation of Gravitational Waves from a Binary Black Hole Merger" \textit{Phys. Rev. Lett.}  \textbf{116} (2016) 061102
  \bibitem{Abbott17} B. P. Abbott \emph{et al.} ``GW170814: A Three-Detector Observation of Gravitational Waves from a Binary Black Hole Coalescence",\textit{Phys. Rev. Lett.}  \textbf{119} (2017)  141101 
\bibitem{Abbott17b} B. P. Abbott \emph{et al.}
`` GW170817: Observation of Gravitational Waves from a Binary Neutron Star Inspiral", \textit{Phys. Rev. Lett.}  \textbf{119} (2017)  161101 
\bibitem{Schumacher23} K. Schumacher, N. Yunes, and K. Yagi, 
“Gravitational wave polarizations with different propagation speeds,” 
Phys. Rev. D \textbf{108}, 104038 (2023).
\bibitem{Schumacher25} K. Schumacher, C. Talbot, D. E. Holz, and N. Yunes, 
“Better early than never: A new test for superluminal gravitational wave polarizations,” 
Phys. Rev. D \textbf{112}, 024067 (2025).
\bibitem{Dong} YQ .Dong,   XB.  Lai, YQ. Liu  et al. ``Gravitational-wave effects in the most general vector–tensor theory". \emph{Eur. Phys. J. C}  85, 645 (2025).
\bibitem{Lai} XB. Lai,  YQ .Dong, YQ. Liu  and YX. Liu, ``Polarization modes of gravitational waves in general Einstein-vector theory", \emph {Phys. Rev. D.} 110.064073 (2024).
\bibitem{Wu} J. Wu and J. Li, `` Prospects of constraining on the polarizations of gravitational waves from binary black holes using space- and ground-based detectors", \emph{Phys. Rev. D.} 110.084057 (2024).
\bibitem{mixed} K. Chatziioannou, N. Yunes, and N. Cornish, ``Morphology-independent tests of mixed polarization content,” \emph{Phys. Rev. D} \textbf{104}, 124067 (2021)
\bibitem{Friedman22}  Y. Friedman, ``A unifying physically meaningful relativistic action". Sci Rep 12, 10843 (2022). https://www.nature.com/articles/s41598-022-14740-7
\bibitem{ana}  Y. Friedman and Scarr \emph{A Novel Approach to Relativistic Dynamics. Integrating Gravity, Electromagnetism and Optics} Fundamental Theories of Physics 210, Springer International Publishing AG  (2023)
\bibitem{FSuper} Y. Friedman, ``Superposition Principle in Relativistic Gravity," \emph{Phys. Scr.}, vol. 99,  105045 (2024)
\bibitem{FStav} Y. Friedman and S. Stav,
 New metrics of a spherically symmetric gravitational field passing classical tests of General Relativity. 
  Europhys. Lett. \textbf{126}, 29001 (2019);  Erratum: Europhys. Lett. \textbf{127}, 19901 (2019) 
\bibitem{FK} Y. Friedman and S. I. Klimovsky ``The Relativistic Gravitational Field of a Spherically Symmetric Extended Body" \emph{Physica Scripta}, 100, 105024,  (2025)   DOI: 10.1088/1402-4896/ae1265  
\bibitem{HellingsDowns1983}
R. W. Hellings and G. S. Downs,
Astrophys. J. Lett. \textbf{265}, L39 (1983)

\bibitem{Lee2008}
K. J. Lee, F. A. Jenet, and R. H. Price,
Astrophys. J. \textbf{685}, 1304 (2008)

\bibitem{Gair2015}
J. Gair, J. D. Romano, and S. R. Taylor,
Phys. Rev. D \textbf{92}, 102003 (2015)

\bibitem{NANOGrav2023}
NANOGrav Collaboration,
Astrophys. J. Lett. \textbf{951}, L8 (2023)

\bibitem{Antoniadis2023}
J. Antoniadis et al. (EPTA Collaboration),
Astron. Astrophys. \textbf{678}, A50 (2023)

\bibitem{Yunes} N. Yunes, X. Siemens, K. Yagi ``Gravitational-wave tests of general relativity with groundbased detectors and pulsar-timing arrays", \emph{Living Reviews in Relativity}  28:3 (2025)

\end{thebibliography}
\end{document}